\documentclass[a4paper,fleqn,usenatbib]{mnras}

\usepackage[T1]{fontenc}
\usepackage{ae,aecompl}

\usepackage{graphicx}	
\usepackage{amsmath}	
\usepackage{amssymb}	

\def\msun{{\rm\,M_\odot}}

\def\gtsima{$\; \buildrel > \over \sim \;$}
\def\simgt{\lower.5ex\hbox{\gtsima}}

\def\msun{\hbox{M$_\odot$}}
\def\songmei2022{https://doi.org/10.48550/arXiv.2212.11034}

\def\inpress{Romano, D. et al., 2023, arXiv:2305.15355}

\title[The Milky Way globular cluster M5]{The dual nature of 
 the tidal tails of NGC~5904 (M5)}

\author[Andr\'es E. Piatti]{
Andr\'es E. Piatti$^{1,2}$\thanks{E-mail: andres.piatti@fcen.uncu.edu.ar} \\
$^{1}$Instituto Interdisciplinario de Ciencias B\'asicas (ICB), CONICET-UNCUYO, 
Padre J. Contreras 1300, M5502JMA, Mendoza, Argentina\\
$^{2}$Consejo Nacional de Investigaciones Cient\'{\i}ficas y T\'ecnicas, Godoy Cruz 
2290, C1425FQB,  Buenos Aires, Argentina\\
}

\date{Accepted XXX. Received YYY; in original form ZZZ}

\pubyear{2022}

\begin{document}
\label{firstpage}
\pagerange{\pageref{firstpage}--\pageref{lastpage}}
\maketitle

\begin{abstract}
The tangential velocity dispersion of stars belonging to the Milky Way globular cluster's tidal tails has 
recently been found  from N-body simulations to be a parameter that distinguishes between
cored and cuspy profiles of low-mass dwarf galaxy dark matter subhaloes where that 
globular cluster formed, and the in-situ formation scenario. In this context, we discovered that M5's 
tidal tails are composed by stars  at two different metallicity regimes ([Fe/H] $\sim$ -1.4 dex 
and -2.0 dex).  The more metal-rich tidal tail stars are of the same metal content than
M5's members and have a tangential velocity dispersion
that coincides with the predicted value for a cuspy formation scenario (subhalo mass 
$\sim$ 10$^9$ $\msun$). The more metal-poor stars, that are
found along the entire M5 tidal tails and have similar distributions to their more metal-rich
counterparts in the M5 colour-magnitude diagram and orbit trajectory, have a
tangential velocity dispersion that refers to a cored  subhalo (mass $\sim$ 10$^9$ $\msun$) 
or an in-situ formation scenario. In order to reconcile the dual distribution of M5 tidal tail stars,
in kinematics and chemistry, we propose that M5 collided with another more metal-poor and
less massive globular cluster anytime before or after it was accreted into the Milky Way.
\end{abstract} 




\section{Introduction}

Some recent detection of tidal tails of Milky Way globular clusters using
clustering search techniques in an N-dimensional phase space assume that 
their stars and those belonging to the cluster have similar proper motions.
In general, an upper limit of 2 mas/yr around the mean cluster
proper motion has been used to identify tidal tails stars, which have been 
called a cold stellar stream \citep{sollima2020,yangetal2022,koposovetal2023}.
However, because of tidal tails stars have speed up their pace in order to escape the 
cluster, their proper motions can be different from the cluster mean 
proper motion. There are, additionally, other reasons that can contribute to make the
space motion of tidal tails stars different from the mean cluster
space velocity, among them, projection effects of the tidal tails, 
the Milky Way tidal interaction, the intrinsic kinematic agitation of a 
stellar stream \citep{wanetal2023}.

Furthermore, some stellar streams have resulted to be kinematically hot, like the
C-19 stream \citep{yuantal2022}, which has been found to be dominated by a dark 
matter halo \citep{erranietal2022}. Rising velocity dispersion profiles
toward the outer regions of globular clusters were also suggested by \citet{cg2021}
for globular clusters placed at the centre of dark matter mini-haloes; a behaviour that 
was also explained by the effects of the  Milky Way tidal interaction \citep{vb2022}.
Recently, \citet{malhanetal2022} showed that globular clusters formed in low-mass
dwarf galaxy dark matter subhaloes, later accreted into the Milky Way, have tidal tails
with a mean tangential velocity dispersion larger than that for tidal tails of globular 
clusters formed in-situ. Based on these outcomes, globular cluster origins
 could be inferred by differentiating whether their tidal tails
are kinematically cold (in-situ formation) or hot (accreted
origin). 

Although some globular clusters do not present tidal tails \citep{zhangetal2022},
it is worth measuring the tangential velocity dispersion of globular cluster
tidal tails in order to distinguish those formed in cored or cuspy cold dark matter subhaloes
(those accreted) from those formed in the Milky Way.
These outcomes can be useful, for instance, to know the nature of the 
dwarf galaxy dark matter subhaloes where the clusters formed, the mass of those
subhaloes, as well as to confirm the previous known associated origins of the 
Milky Way globular clusters \cite[e.g.][]{massarietal2019}. We embarked
 in this challenging analysis by examining the tidal tails of NGC~5904 (M5), a globular
 clusters associated to one of the latest merger events occurred in the Milky Way 
 \citep{kruijssenetal2019,forbes2020}.

In this Letter we report the discovery of the dual nature of the tidal tails of M5. They
contain stars both in the cored and cuspy formation scenarios, 
which in turn have clearly different overall metallicity contents. In Section 2 we
present the analysis of the data, while in Section 3 we  speculate on a possible scenario 
for the resulting outcomes.

\section{Data analysis}

\citet{g2019} detected using the second release of the {\it Gaia} database 
\citep{gaiaetal2016,gaiaetal2022b} a long trailing tidal tail extending westward from M5. He 
selected 50 highest-ranked tidal tail member candidates based on their similar 
distances, their magnitudes and colours distributed along the M5 colour-magnitude
diagram, their proper motions consistent with the cluster's trajectory at a
detection significance $\approx$ 10$\sigma$. The long tidal tail is included in the
recent atlas of Milky Way streams compiled by \citet{mateu2023}. We used the
 derived  {\it Gaia} DR3 parameters\footnote{Kindly provided by Cecilia Mateu.} for 
these 50 stars to compute their tangential velocities.

The top-left panel of Fig.~\ref{fig1} shows the distribution of the stars in the sky. 
As for their physical distances, we relied on the results found by \citet{g2019} and
\citet{ibataetal2021}, who place the long tidal tail at a constant distant, that of the
M5's mean heliocentric distance \citep[7.48 kpc;][]{bv2021}, as is illustrated by the solid line in the
parallax versus R.A.plot drawn in the top-right panel of Fig.~\ref{fig1}. Hence, we 
adopted for the subsequent analysis the R.A. coordinates as the tidal tail tracing 
coordinates. 

The tangential velocities were computed as $V_{Tan}$ = $k \times d_\odot \times \mu$;
where $k$ = 4.7405 km s$^{\rm -1}$ kpc$^{\rm -1}$ (mas/yr)$^{\rm -1}$, $d_\odot$ is the
tidal tail distance, and $\mu = \sqrt{{\mu_{\alpha}^*}^2 + \mu_{\delta}^2}$, with
$\mu_{\alpha}^*$ and $\mu_{\delta}$ being the proper motions in R.A. and Dec as
provided by  {\it Gaia} DR3.  We used Figure~15 of \citet{ibataetal2021} to
estimate the mean $d_\odot$ for intervals of $\Delta$(R.A.) = 5$\degr$, and used those values
to compute  $V_{Tan}$ for stars in the respective R.A. bins.
The bottom-right panel of Fig.~\ref{fig1} shows the
observed relation between $V_{Tan}$ and R.A. The error bars come from
propagation of errors of the $V_{Tan}$ expression. We then fitted a second order
polynomial function, represented by the solid line in the bottom-right panel of
Fig.~\ref{fig1}, and computed the different between the measured $V_{Tan}$
values and the corresponding ones on the fitted function for the respective 
R.A.
 
In order to obtain the tangential velocity dispersion,
we derived the dispersion of the resulting residual distribution 
by employing a maximum likelihood approach \citep[see, e.g.,][]{pm1993,walkeretal2006}. 
For that purpose, we optimized the probability $\mathcal{L}$  given by : \\

\noindent $\small
\mathcal{L}\,=\,\prod_{i=1}^N\,\left( \, 2\pi\,(\sigma_i^2 + W^2 \, ) 
\right)^{-\frac{1}{2}}\,\exp \left(-\frac{(\Delta(V_{Tan})_i \,- <\Delta(V_{Tan})>)^2}{2(\sigma_i^2 + W^2)}
\right) \\$

\noindent where $\Delta(V_{Tan})_i$ and $\sigma_i$ are the residiual $V_{Tan}$ value and 
the corresponding error for the $i$-th star. We obtained a mean tangential velocity 
dispersion $W$ =  15.65 $\pm$ 0.47 km/s. This result largely exceeds the highest predicted
tangential velocity dispersion for  globular cluster streams in dark
matter subhaloes with a mass of 10$^9$ $\msun$ ($\sim$ 8.5 km/s). 

We used the overall metallicity estimates ([Fe/H]) and their uncertainties
provided by GSP-Phot in {\it Gaia} DR3 to
check whether the resulting $W$ value can be biased by the presence of field stars. 
Only 25 out of the 50 stars have available {\it Gaia} DR3 metallicities. For them, we first corrected 
the [Fe/H] values following the prescriptions given by  \citet{andraeetal2022}\footnote{https://www.cosmos.esa.int/web/gaia/dr3-gspphot-metallicity-calibration} and then plotted the
resulting values as a function of R.A. (see bottom-left panel of Fig.~\ref{fig1}). As can be seen, 
there are two different metallicity regimes, centred at [Fe/H] $\sim$ -1.4 dex and -2.0 dex,
respectively.  For each of them, we repeated the above procedure to compute the
tangential velocity dispersion, and obtained $W$ = 7.50 $\pm$ 1.38 km/s and
2.00 $\pm$ 4.07 km/s, for the more metal-rich and more metal-poor samples, respectively.
For the most metal-rich regime, we did not consider the star at [Fe/H] $\approx$ -0.5 dex,
 because it is beyond the mean value by more than 7 times the metallicity dispersion.
We finally searched the {\it Gaia} DR3 database looking for M5 members with 
metallcity estimates,  with the aim of validating the above procedure and results. 
We applied the selection cuts as in \citet{vb2021}, selecting stars with
$\mu_\alpha^*$ and $\mu_\delta$ values within the dispersion found by 
\citet{baumgardtetal2019}\footnote{https://people.smp.uq.edu.au/HolgerBaumgardt/globular/},
\texttt{ruwe} $<$ 1.15, \texttt{visibility\_periods\_used}  $\le$ 10, and \texttt{ipd\_frac\_multi\_peak} 
$\le$ 2. We found four stars within the cluster area ($<$ 5$\arcmin$) with  corrected [Fe/H]
values of $\sim$ -1.4 dex, which is in excellent agreement with the known M5 metal content
\citep[{\rm [Fe/H]} = -1.33;][and references therein]{vdgetal2022}. 

\section{Discussion and conclusions}

M5, with an age of 11.46$\pm$0.44 Gyr \citep{vdgetal2022}, is associated to the Helmi stream, 
 the aftermath of a merger event (5-8 Gyr) of a small mass dwarf galaxy experienced by the Milky Way 
\citep{massarietal2019,kruijssenetal2019,forbes2020}. Its orbit eccentricity (0.79$\pm$0.01) and 
inclination (74.09$\pm$0.66$\degr$) also witness its accreted origin \citep{piattietal2019b}.
 If we entered in the bottom panel of Figure 1 of \citet{malhanetal2022} with $W$ = 7.5 km/s - the 
present tangential velocity dispersion for M5 tidal tail stars with similar cluster metallicities  -, we
would find that the cluster formed in a dwarf galaxy dark matter
subhalo with a cuspy profile with a mass of $\sim$  10$^{8.8\pm0.1}$ $\msun$.
Therefore, we can conclude that this component of the stream is consistent with that formation scenario.

\begin{figure*}
\includegraphics[width=\textwidth]{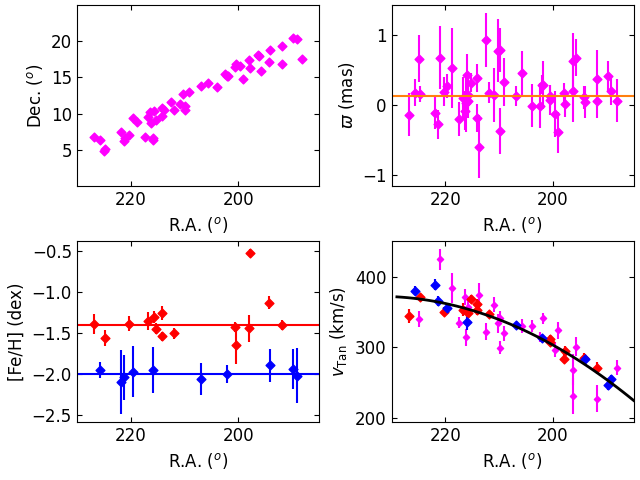}
\caption{Highest-ranked candidate tidal tail stars according to \citet{g2019}. Data were
taken from \citet{mateu2023}.  The
orange line in the upper-right panel represents the mean cluster parallax derived by
\citet{bv2021}. Mean {\it Gaia} metallicities and uncertainties are corrected according to \citet{andraeetal2022}; the red and blue lines represent the mean values of red and
blue points, respectively. A quadratically least-square fit is shown in the $v_{Tan}$ versus
R.A. plane for all the data, included those without metallicities (magenta points).}
\label{fig1}
\end{figure*}

However, the dual metallicity distribution found among the highest-ranked tidal tail candidate
members hampers our smooth understanding of the M5's origin. Stars at both metallicity regimes
are found distributed along the entire extension of the examined tidal tail (see Fig.~\ref{fig1}), 
which undoubtedly removes
any speculation that most of them are unrelated stars to M5. On the other hand, overall 
metallicity differences $\Delta$[Fe/H] larger than $\sim$ 0.6 dex  (see Fig.~\ref{fig1}) were  found in the 
building block globular clusters Terzan~5 \citet{romanoetal2023} and Liller~1 \citep{crociatietal2023}
in the Milky Way and M54 \citep{alfarocuelloeal2019} in the Sagittarius dwarf spheroidal galaxy. 
 We note that the
difference in metallicity between M5 1G and 2G stars is 0.03 dex \citep{marinoetal2019}. Hence,
M5 would not seem to belong to that globular cluster group, since it does
not show the expected age spread  (see its $HST$ colour-magnitud diagram in \citet{vdgetal2022})
for a long star formation 
history as observed in those
building block globular clusters.  Therefore, we can conclude that only tidal tail stars with a metallicity 
similar to that of M5 were born in the cluster itself. Because of the remarkable different metallicity, and
according to the known nucleosyhthesis processes, the more 
metal-poor tidal tail stars could not form in M5.

The mean tangential velocity dispersion of the more metal-poor stars ($W$ = 2.00) matches the
cored dark matter subhalo formation scenario (mass $\sim$ 10$^9$ $\msun$) in the N-body 
simulation performed by \citet{malhanetal2022}. If this were the case, then these more
metal-poor stars could be the relicts of a globular cluster formed in another dwarf galaxy,
that collided with M5.  Note that the lack of metallicity spread among them (see Fig.~\ref{fig1}) discards any possible accreation of stellar structures with a long star formation
history or scaterred field stars. The collision could take place before M5 was accreted into the Milky
 Way or in the Milky Way itself, after they were accreted into the Galaxy, separately. Because of the
 uncertainty in the resulting $W$ value ($\sigma$$W$ = 4.07 km/s) the more metal-poor
 stars could also be the fossils of a globular cluster formed in-situ that collided with M5.
Whatever the scenarios is considered, it would seem that a reasonable interpretation 
 for the dual nature of stars belonging to the tidal tails of M5,  in kinematics and chemistry,  is that
 the cluster experienced an encounter with another globular cluster.
 
 This speculation opens this research field to further analyses. Indeed, a spectroscopic survey
 of tidal tails stars in M5 is mandatory, as well as in M5's main body. According to the number of
 analysed tidal tail stars with metallicity estimates, the putative colliding cluster could contain $\sim$ 
 1/3 of the M5's tidal tail mass. We also wonder whether some stars of such a 
 disrupting cluster could be trapped within the M5's main body. Alternatively, future N-body simulations 
 will be very useful to analyse this scenario in detail. \citet{ew2022} simulated a dwarf galaxy passing 
 within the M5's tidal tails and  found that it is negligibly perturbed by the dwarf galaxy. 
In case future simulations find that the collisional scenario is feasible,  then that would further imply that observations of metallicity spread in M5 do not necessarily require multiple episodes of star formation, but that it can occur as a result of collision between two globular clusters hosting different stellar populations. 
The speculated collision between M5 and another more metal-poor globular cluster is not the prototypical phenomenon proposed by theories of the formation of globular clusters, however, its discovery is both encouraging and interesting. Therefore, M5 provides a unique laboratory to explore various new aspects 
of formation and evolution theory of globular clusters.

\section*{Acknowledgements}
We thank the referee for the thorough reading of the manuscript and timely 
suggestions to improve it.
This work has made use of data from the European Space Agency (ESA) mission
{\it Gaia} (\url{https://www.cosmos.esa.int/gaia}), processed by the {\it Gaia}
Data Processing and Analysis Consortium (DPAC,
\url{https://www.cosmos.esa.int/web/gaia/dpac/consortium}). Funding for the DPAC
has been provided by national institutions, in particular the institutions
participating in the {\it Gaia} Multilateral Agreement.


\section{Data availability}

Data used in this work are available upon request to the author.







\bsp	
\label{lastpage}
\end{document}